\documentclass[preprint,amsmath,eqsecnum,aps,floats,groupedaddress]{revtex4}

\usepackage{graphicx}% Include figure files
\usepackage{dcolumn}% Align table columns on decimal point
\usepackage{bm}% bold math


\begin{document}

\title{Energy gaps in the failed high-T$_c$ superconductor La$_{1.875}$Ba$_{0.125}$CuO$_4$}

\author{Rui-Hua He$^1$, Kiyohisa Tanaka$^{1, 2, 6}$, Sung-Kwan Mo$^{1, 2}$,
Takao Sasagawa$^{1, 4}$, Masaki Fujita$^3$, Tadashi Adachi$^5$,
Norman Mannella$^{1, 2}$\footnote{Present address: Department of
Physics and Astronomy, University of Tennessee, Knoxville, Tennessee
37996, USA}, Kazuyoshi Yamada$^3$, Yoji Koike$^5$, Zahid
Hussain$^2$, Zhi-Xun Shen$^1$}

\affiliation{$^1$Department of Physics, Applied Physics and
Stanford Synchrotron Radiation Laboratory, Stanford University,
Stanford, California 94305, USA}

\affiliation{$^2$Advanced Light Source, Lawrence Berkeley National
Lab, Berkeley, California 94720, USA}

\affiliation{$^3$Institute of Materials Research, Tohoku
University, Sendai 980-8577, Japan}

\affiliation{$^4$Materials and Structures Laboratory, Tokyo
Institute of Technology, Kanagawa 226-8503, Japan}

\affiliation{$^5$Department of Applied Physics, Tohoku University,
Sendai 980-8579, Japan}

\affiliation{$^6$Department of Physics, Osaka University, Osaka
560-0043, Japan}

\date{February 14, 2008}

\maketitle

\begin{raggedright}
\parindent=0.5in

%\begin{abstract}
\textbf{A central issue on high-T$_c$ superconductivity is the
nature of the normal-state gap (pseudogap)
\cite{HTSC:PeudogapReview} in the underdoped regime and its
relationship with superconductivity. Despite persistent efforts,
theoretical ideas for the pseudogap evolve around fluctuating
superconductivity \cite{HTSC:PrecursorPairing}, competing order
\cite{HTSC:CompetingOrders_Staggeredflux,
HTSC:CompetingOrders_DDW, HTSC:CompetingOrders_VarmaLoops,
HTSC:CompetingOrders_CDW_DiCastro, HTSC:CompetingOrders_CDW_Lee, HTSC:CompetingOrders_SDW_Markiewicz}
and spectral weight suppression due to many-body effects
\cite{e-ph:polaron:t-JPolaron}. Recently, while some experiments
in the superconducting state indicate a distinction between the
superconducting gap and pseudogap
\cite{cuprates:Bi2212:Kiyo_TwoGap, cuprates:Bi2212:WeiShengUDBCS,
cuprates:Bi2201:Kondo_TwoGap, cuprates:LSCO:Takahashi_TwoGap,
cuprates:Bi2201:STMTwoGap_Hudson}, others in the normal state,
either by extrapolation from high-temperature data
\cite{cuprates:Bi2212:NodalMetal} or directly from
La$_{1.875}$Ba$_{0.125}$CuO$_4$ (LBCO-1/8) at low temperature
\cite{stripe:other:Valla}, suggest the ground-state pseudogap is a
single gap of $d$-wave \cite{HTSC:dwaveSC} form. Here we report
angle-resolved photoemission (ARPES) data from LBCO-1/8, collected
with improved experimental conditions, that reveal the
ground-state pseudogap has a pronounced deviation from the simple
$d$-wave form. It contains two distinct components: a $d$-wave
component within an extended region around the node and the other
abruptly enhanced close to the antinode, pointing to a dual nature
of the pseudogap in this failed high-T$_c$ superconductor which
involves a possible precursor pairing energy scale around the node and
another of different but unknown origin near the
antinode.}

%\end{abstract}
%\pacs{}

The first high-T$_c$ superconductor discovered,
La$_{2-x}$Ba$_{x}$CuO$_4$ (LBCO), holds a unique position in the
field because of an anomalously strong bulk T$_c$ suppression near $x=1/8$. Right around this "magic" doping level, scattering experiments by neutron \cite{stripe:neutron:LBCO_Tranquada, stripe:neutron:LBCO_Fujita}
and X-ray \cite{stripe:other:LBCO_Xray} find a static spin and
charge (stripe) order. By itself, this observation raises a series
of intriguing questions: whether the stripe order is a competing
order that suppresses the superconductivity in LBCO-1/8 and, if
the answer is positive, which aspect, the pairing strength or the
phase coherence, is involved in the T$_c$ suppression and how this
mechanism applies to other dopings or families. For our
investigation of the "ground-state" pseudogap, as defined in Ref.
\cite{stripe:other:Valla} that ignores the residual
superconductivity, its sufficiently high doping yet extremely low bulk T$_c$ ($\sim$ 4 K) makes LBCO-1/8 an ideal system: especially for the small-gap measurement near the node, difficulties due to either the
unscreened disorder potential, a problem for extremely low doping,
or trivial thermal broadening, a problem above T$_c$ for higher
doping, are circumvented.

While thermal effects require an extrapolation from
high-temperature data to obtain the ground state physics
\cite{cuprates:Bi2212:NodalMetal}, a direct measurement on
LBCO-1/8 at low temperature has been made
\cite{stripe:other:Valla}. With experimental resolutions
compromised to obtain sufficient signal to noise, a simple
$d$-wave gap function is reported with no discernible nodal
quasi-particles found. Given the importance of this issue, we have
performed an ARPES \cite{ZXreview} study of LBCO-1/8 at T$>$T$_c$ with much improved resolutions in a measurement geometry favorable for the
detection of nodal quasi-particles (see Fig. \ref{Fig. 1} \&
Supplementary Section I). Our data provide two important new
insights. First, there is a well-defined nodal quasi-particle peak
suggesting nodal quasi-particles can exist in the stripe ordered
state. Second, there is a rich gap structure suggesting the
pseudogap physics is more elaborate than the simple $d$-wave
version. In particular, a new kind of pseudogap, which is not
smoothly connected to the usual one tied to the antinodal region,
can exist in the nodal region when superconductivity is suppressed
due to the loss of phase coherence.

As shown in Fig. \ref{Fig. 1}, there exists a well-defined quasi-particle peak in the energy distribution curves (EDC's) at the Fermi crossing points (k$_F$'s) around the node. Upon dispersion towards the antinode, the lineshape quickly becomes incoherent. Data taken with different photon energies (h$\nu$'s) in different Brillouin zones (BZ's) show consistent results (Supplementary Fig. S1), providing the first unambiguous piece of direct evidence that nodal quasi-particles survive in the stripe ordered state. This suggests these two seemingly very different aspects of cuprate phenomenology can be compatible with each other
\cite{stripe:theory:nodalQPvsStripe}.

The observation of quasi-particle peaks in the EDC's gives a firm
foundation for the gap analysis around the node. However, given
the small gap size as well as a relatively small peak-background
intensity ratio and large quasi-particle peak linewidth ($\Gamma$)
(Fig. \ref{Fig. 1}d), quantitative determination of the gap is
non-trivial in LBCO-1/8. In the following, we will present
analysis that addresses two important aspects of our data: i) the
gap measured by the leading edge gap (LEG) method, same as
employed by Ref. \cite{stripe:other:Valla}, is shown to have two
components and cannot fit a simple $d$-wave form (Figs. \ref{Fig. 2} \& \ref{Fig. 4}); ii) the gap around the node, measured either
by the LEG method (Fig. \ref{Fig. 2}) or by a curve fitting
procedure commonly used in the field (Fig. \ref{Fig. 3}), is shown
to be $d$-wave like with a finite gap slope.

In Fig. \ref{Fig. 2}a, from the systematic shift of the
leading-edge midpoint along the energy axis, we notice the LEG
keeps increasing from the node towards the antinode with the rate of
increase getting larger close to the antinode. This is more
clearly shown in Fig. \ref{Fig. 2}b where the momentum dependence
of the extracted LEG from two selected samples is plotted.
Consistent results are obtained at different times after sample
cleaving, showing no signs of sample aging which could affect
small-gap measurements, and in different BZ's with different
h$\nu$'s on another sample. They unanimously point to a pronounced
deviation of the gap function from the simple $d$-wave form. As
detailed in Supplementary Section IIIB, C\& IV, this observation
goes beyond the interpretation based on either a simple $d$-wave
gap under the influence of experimental resolutions (Supplementary
Fig. S10) and the lineshape decoherence towards the antinode
(Supplementary Fig. S11) or a single pairing gap with the
inclusion of higher $d$-wave gap harmonics
\cite{cuprates:Bi2212:dwaveHarmonics} (Supplementary Fig. S7). It
naturally reveals a striking characteristic: this normal-state gap
has two distinct components, with the one around the node (the
nodal gap) exhibiting a simple $d$-wave form, i.e., a linear
dependence with respect to (w.r.t.) $[cos(k_x)-cos(k_y)]/2$,
along the underlying FS over a significant momentum range and the
other setting in near the antinode (the antinodal gap) which
deviates sharply from its nodal counterpart.

Although it is clear that the LEG opens in a $d$-wave fashion
around the node (inset of Fig. \ref{Fig. 2}a), it is too crude to
conclude the real gap ($\Delta$) function of the system is also
simple $d$-wave like because the leading edge shift in principle
can be due to the variation in $\Gamma$ even if $\Delta$ is fixed
\cite{misc:PES:LEGsim}. This alternative has to be explored
especially for LBCO-1/8 where $\Gamma>>\Delta$ near the node.
Thus, we fit the E$_F$-symmetrized EDC's at k$_F$ to a
phenomenological model \cite{cuprates:Bi2212:Fitting_NormanModel}
which assumes a self energy,
$\Sigma(k_F,\omega)=-i\Gamma+\frac{\Delta^2}{\omega}$ (Fit Model),
where $\Gamma$ and $\Delta$ are subject to the fit (Fig. \ref{Fig. 3}a). Because of the small peak-background ratio, the fitting
results are somewhat affected by the background subtraction. In
Fig. \ref{Fig. 3}b, we show the results of $\Delta$ without any
background subtraction, or with an MDC constant background or an
EDC integral background subtraction (Supplementary Fig. S5). Two
general trends of the results are: i) regardless of methods for
background subtraction, an initial gap slope is robustly defined
close to the node where quasi-particle peaks are clearly present
in e1 $\sim$ e3; ii) upon weakening of the peak feature, a large
background dependence appears starting from e4. For e4, with more
background subtracted from the high binding energy side of the
peak feature, $\Delta$ decreases and tends to fall onto the
initial gap slope. For the
completeness of our analysis, we have also shown in Supplementary
Fig. S6 fitting results based on another model that fits the data less well. While the quantitative gap values are model dependent, its $d$-wave form remains robust (Supplementary Section IID).

Summarizing the above, both the model independent (the LEG
method) and model dependent (the curve fitting procedure) gap
analysis suggest that the nodal gap opens in a $d$-wave fashion with a finite slope. Despite its presence in the normal state, it is highly suggestive that this nodal gap is of a pairing origin based on the following reasons: i) its $d$-wave form is consistent of the generic pairing symmetry of cuprate superconductors, particularly in La-based cuprates, where a similar gap in the nodal region is found to be related to superconductivity \cite{cuprates:Bi2201:Kondo_TwoGap}; ii) its gap slope strikingly coincides within error bars with those of La$_{2-x}$Sr$_{x}$CuO$_4$ (LSCO) $x=0.11$ \cite{cuprates:LSCO:TakaoLSCOGrowth_2} and LBCO $x=0.083$
\cite{stripe:other:LBCO_Adachi} (T$_c$=26 K and 23 K, respectively, see Supplementary Fig. S8 for raw data), which are both in the superconducting state (Fig. \ref{Fig. 4}a). Compared with the one analyzed using Fit Model in Bi2212 of a similar doping level (UD75K in Ref. \cite{cuprates:Bi2212:WeiShengUDBCS}) at low temperature, it exhibits a factor of $\sim$ 2 reduction, reminiscent of the optimal T$_c$ difference of these two families; iii) its susceptibility to thermal smearing in contrast to its antinodal counterpart (Fig. \ref{Fig. 4}b) is reminiscent of the cases in other superconducting cuprates \cite{cuprates:Bi2212:WeiShengUDBCS, cuprates:Bi2201:Kondo_TwoGap}. The existence of precursor pairing in the normal state of LBCO-1/8 is further supported by recent transport measurements \cite{stripe:other:Tranquada2DSC}. A drop in the in-plane resistivity at T$_{2D}\sim 40$ K, where the concurrent stripe (as a density wave) formation would often result in an increase of resistivity, implies an onset of superconducting fluctuations. Although the coincidence of its partial closing with the simultaneous resistivity increase at T=T$_{2D}$ upon heating (Fig. \ref{Fig. 4}b) suggests the precursor pairing origin of the nodal gap rather than the conventional density wave type, understanding the relationship between these two coexisting orders in the system at T$<$T$_{2D}$ still poses a challenge.

On the other hand, the pairing strength, as reflected by the slope of the nodal pairing gap, is comparable in LBCO-1/8 and its neighboring compounds of much higher bulk T$_c$ values (Fig. \ref{Fig. 4}a). Hence, a natural explanation for the bulk T$_c$ suppression in LBCO-1/8 is its lack of a global superconducting phase coherence. Interestingly, this loss of superconducting phase coherence coincides with the stabilization of the stripe order at x$\sim$1/8. Intrigued by this, theorists have proposed that the global superconducting phase coherence can be prohibited via the dynamical interlayer decoupling in the system where superconductivity is modulated by the stripe order of some specific configurations \cite{stripe:theory:DynamicDecoupling_1st,
stripe:theory:DynamicDecoupling_Steve}. Nevertheless, the modulated superconductivity of a non-zero wave vector generally does not produce a $d$-wave gap with a point node as observed here. Our finding would put a strong constraint on future theoretical attempts to resolve the microscopic mechanism for the failure of high-T$_c$ superconductivity in LBCO-1/8.

As suggested by its different momentum dependence from its nodal counterpart, the antinodal pseudogap may have a different origin, which has been the
subject for intense discussions in literature \cite{HTSC:PrecursorPairing, HTSC:CompetingOrders_Staggeredflux,
HTSC:CompetingOrders_DDW, HTSC:CompetingOrders_VarmaLoops,
HTSC:CompetingOrders_CDW_DiCastro, HTSC:CompetingOrders_CDW_Lee, HTSC:CompetingOrders_SDW_Markiewicz, e-ph:polaron:t-JPolaron}. Generally, as inferred from Supplementary Fig. S11, a similar antinodal phenomenology as we observed can be achieved either by i) a true gap opening together with a strong quasi-particle scattering (the $\Delta-\Gamma$ physics) or by ii) a great suppression of the quasi-particle spectral weight (the Z physics). While attempts of Scheme i) working in the weak coupling approach can produce a qualitatively similar quasi-particle gap structure by considering the competition between superconductivity and the charge \cite{HTSC:CompetingOrders_CDW_DiCastro, HTSC:CompetingOrders_CDW_Lee} or/and spin \cite{HTSC:CompetingOrders_SDW_Markiewicz, stripe:theory:DynamicDecoupling_Steve}
density wave order, Scheme ii) demands a strong-coupling route (e.g., Ref. \cite{e-ph:polaron:t-JPolaron}) where the extended quasi-particle analysis as presented above might break down due to its incapability of capturing the lowest-lying excitations of a vanishing spectral weight. In any case, new physics other than the nodal-type precursor pairing alone is required to fully capture the essence of the antinodal pseudogap. 

In contrast to the notion of a simple $d$-wave nodal liquid as the pseudogap ground state which is directly derived from the antinodal
pseudogap \cite{cuprates:Bi2212:NodalMetal, stripe:other:Valla},
our observation of an apparent break-up of the gap function
suggests a very different picture. It reveals a much richer
pseudogap physics with its two aspects manifesting differently in
distinct momentum regions, i.e., the nodal precursor pairing and
the antinodal pseudogap of different but unknown origin. These two
aspects might be emphasized differently by different experimental
probes in the normal state, which has led to the two extremes of
ideas for the pseudogap, in particular, whether it has a direct
relationship with pairing. Our results suggest a plausible
reconciliation between them.

\begin{acknowledgments}
The authors would like to acknowledge helpful discussions with H.
Yao, W.-S. Lee, E. Berg, T. P. Devereaux, S. A. Kivelson, X. J. Zhou, T. Xiang and revision of the manuscript by R. G. Moore. R.-H.H. thanks G. Yu for kind assistance in the sample preparation and the SGF for financial supports. The work at the ALS is supported by the DOE Office of Basic Energy Science under Contract No. DE-AC02-05CH11231. This work at Stanford is supported by the DOE Office of Science, Division of Materials Science and Engineering, under Contract No. DE-FG03-01ER45929-A001 and a NSF grant DMR-0604701.
\end{acknowledgments}

\section{Author contributions}
TS, MF and KY, TA and YK provided and prepared the LSCO $x=0.11$, LBCO-1/8 and LBCO $x=0.083$ samples, respectively. SKM, KT and NM maintained the experimental endstation and ensured its high performance. RHH and KT carried out the experiment with the assistance of SKM, NM and TS. RHH did and KT repeated the data analysis and simulations. RHH wrote the paper with helpful suggestions and comments by SKM. ZH and ZXS are responsible for project direction, planning and infrastructure.

\section{Author Information}
Reprints and permission information is available online at
http://npg.nature.com/reprintsandpermissions/. The authors declare
no competing financial interests. Correspondence and requests for
materials should be addressed to Z.-X. Shen (zxshen@stanford.edu).

Supplementary Informatin accompanies this paper on
www.nature.com/naturephysics.

%\newpage
%\References

%%%%%%%%% figures in the main body %%%%%%%%%
\newpage

\begin{figure}
\linespread{1}
\par
%\begin{center}
\hspace*{-0.5cm}
%\end{center}
\par
\linespread{1.6} \caption[LBCO-1/8 FS \& spectra with 55eV]
{\textbf{ARPES spectra and the Fermi surface (FS) of LBCO-1/8.}
All data were taken on Sample A with $h\nu$=55 eV at T=21$\pm$ 2 K
(see Supplementary Fig. S1 for data on Sample B with $h\nu$=110
eV). \textbf{a}, A coarse FS map having a large momentum coverage.
\textbf{b}, A fine FS map taken within 2 $\sim$ 5 hrs after sample
cleaving, covering the yellow-shaded region mostly in the
second BZ for the detailed spectral analysis. Cuts
were taken parallel to the zone diagonal (red arrow) with the
polarization of light fixed in plane and orthogonal to the zone
diagonal (blue arrow). Each dot of the dotted lines corresponds to
an actual sampling momentum position by that cut. The red curves
in \textbf{a} and blue curves in \textbf{b} represent the same
tight-binding FS resulting from a global fit to the data
(Supplementary Fig. S2). \textbf{c}, The momentum distribution
curves (MDC's) at E$_F$, m1 $\sim$ m10, from cuts \#1 $\sim$ \#10
from the nodal to the antinodal as shown in \textbf{b}. k$_F$'s
determined are indicated by red dots (Supplementary Section IIA \&
Fig. S9). Note that k$_F$ on cut \#10 is already past the zone
boundary. The parallel momentum coverage of each MDC does not
correspond to the length of that cut shown in \textbf{b}. The
additional peak features in m9 \& m10 are due to the FS crossings
in the adjacent quadrant of the BZ as shown in \textbf{a}.
\textbf{d}, The EDC's at k$_F$, e1 $\sim$ e10, from cuts \#1
$\sim$ \#10, respectively. Red, blue and green arrows indicate the
peak intensity, background and peak linewidth in e1, respectively.
All spectra are offset vertically for clarity. Labels e1, e4, e7 \& e9 in \textbf{b} denote the momentum positions of the corresponding EDC's. See Supplementary Section I for additional information.} \label{Fig. 1}
\end{figure}

\begin{figure}
\linespread{1}
\par
%\begin{center}
\hspace*{-0.5cm}
%\end{center}
\par
\linespread{1.6} \caption[Two gaps by LEG analysis] {\textbf{The
pseudogap function of LBCO-1/8 by the LEG analysis.} \textbf{a}, Selected
EDC's at k$_F$, e1, e3, e5, e7 \& e8, reproduced from Fig.
\ref{Fig. 1}d in an expanded energy scale which are normalized in
intensity at the leading edge midpoint (LEM)(Supplementary Fig. S3 \& S4). Inset: e1 $\sim$ e4 similarly normalized and shown in a more
expanded energy scale to visualize the LEG opening near the node.
\textbf{b}, LEG plotted as a function of
$[cos(k_x)-cos(k_y)]/2$, for LBCO-1/8 Sample A measured at
different times after sample cleaving with h$\nu$=55 eV; for
Sample B measured in different BZ's with h$\nu$=55 or 110 eV.
T=21$\pm$ 2 K. The results regarding the LEG function at T$\sim 20$
K were repeated on Sample C (Fig. \ref{Fig. 4}b) and another 6
samples from different batches of growth (not shown). Labels e1, e4, e7 \& e9 denote the momentum positions of the corresponding EDC's in Fig. \ref{Fig. 1}d. The black line is a guide to the eye for the ground-state pseudogap function.  The LEG values for cuts at different momentum positions for each sample are referenced to the value from the nodal reference cut taken right after that cut. Error bars are determined by the uncertainty of E$_F$, k$_F$ and the energy window dependence of the LEG by the LEG analysis (see Supplementary Section IIB).} \label{Fig. 2}
\end{figure}

\begin{figure}
\linespread{1}
\par
%\begin{center}
\hspace*{-0.5cm}
%\end{center}
\par
\linespread{1.6} \caption[$d$-wave nodal gap] {\textbf{The nodal
gap analysis by curve fitting.} \textbf{a}, The
E$_F$-symmetrized EDC's near the node, e1 $\sim$ e4, and fit
curves by Fit Model. The arrows are very rough guides to the eye
for the centroid of possible peak features in the EDC's.
\textbf{b}, $\Delta$ from different fits without/with different
methods of background subtraction (Supplementary Section IIC). Labels e1$\sim$e4 denote the momentum positions of the corresponding EDC's in \textbf{a}. The dashed lines are guides to the eye showing the gap slopes by fitting using the two models, respectively. Error bars are
determined by the uncertainty of E$_F$, k$_F$ and the statistical
errors of $\Delta$ from the fits (with its fitting energy window
dependence) but do not include the dependence of background
subtraction (see Supplementary Section IIB). A resolution Gaussian
with $\Delta E=18$ meV is used for the convolution with the
spectral function for the fitting.} \label{Fig. 3}
\end{figure}

\begin{figure}
\linespread{1}
\par
%\begin{center}
\hspace*{-0.5cm}
%\end{center}
\par
\linespread{1.6} \caption[The doping and T dependence]
{\textbf{The doping and temperature dependence of the LEG function.}
\textbf{a}, Comparison of the LEG function between LBCO-1/8 (Sample A, within 2 $\sim$ 5 hrs after sample cleaving, reproduced from Fig. \ref{Fig. 2}b), LSCO $x=0.11$ at T=21$\pm$ 2 K and LBCO $x=0.083$ at T=19$\pm$ 2 K. The dashed line is an eye guide for the antinodal gap component of LBCO $x=0.083$. Inset: Comparison between LBCO-1/8 and LBCO $x=0.083$ of the near-E$_F$ portion of antinodal EDC's taken at the momentum position roughly indicated by the arrow in \textbf{a}, showing the absence of the reported anomaly at x$\sim$1/8 in the antinodal pseudogap size \cite{stripe:other:Valla}. EDC's are normalized in intensity at the LEM and shifted in energy w.r.t. the nodal LEM of each sample. \textbf{b}, The LEG function of LBCO-1/8 (Sample C) at T=19$\pm$ 2 K is compared with the one at T=61$\pm$ 2 K. The dashed green curve is a guide to the eye for the 61 K data. Note that the rapid smearing of the distinction between the two gaps as temperature increases is not captured by the extrapolation scheme used in Ref. \cite{cuprates:Bi2212:NodalMetal} to obtain the ground-state information based on results at high temperatures. Inset: Detailed
temperature dependence in the nodal gap region of Sample A at
three selected momentum positions, C1$\sim$C3, as indicated by
arrows. Note that C3 is close to the cross-over position of the
two gap components. Solid and dashed lines are guides to the eye.
See Supplementary Section IIIA for discussion. The same guidelines in black in \textbf{a} \& \textbf{b} for LBCO-1/8 at low temperature are reproduced from Fig. \ref{Fig. 2}b. All data were taken with h$\nu$=55 eV. The same nodal referencing scheme is used for the LEG values as in Fig. \ref{Fig. 2}b. Error bars are determined by the uncertainty of E$_F$, k$_F$ and the energy window dependence of the LEG by the LEG analysis (see Supplementary Section IIB).} \label{Fig. 4}
\end{figure}

\clearpage
\bigskip
\begin{figure}
\hspace*{-1cm}
\includegraphics [width=7in]{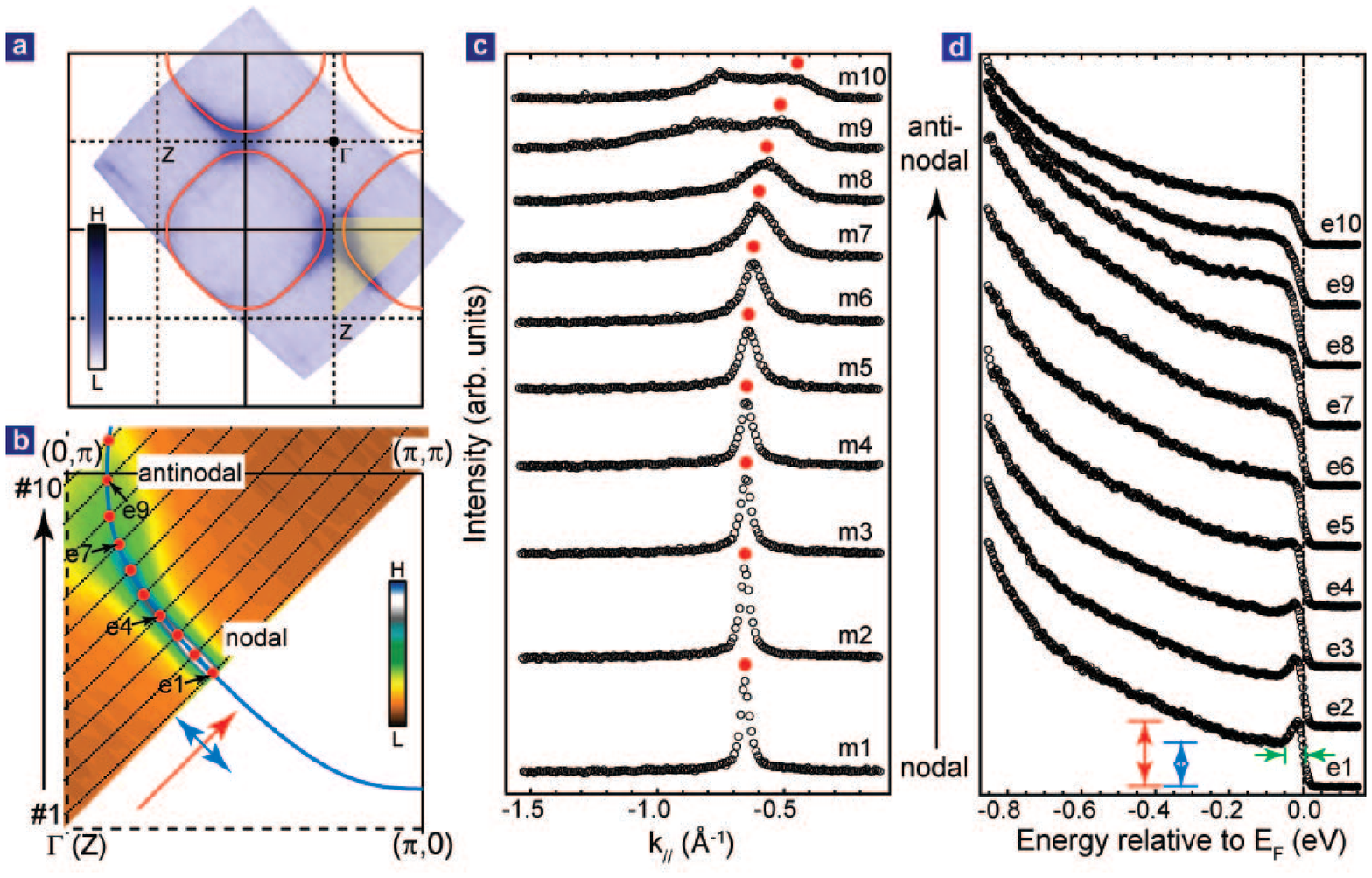}

Figure \ref{Fig. 1}
\end{figure}

\clearpage
\bigskip
\begin{figure}
\hspace*{-1cm}
\includegraphics [width=4 in]{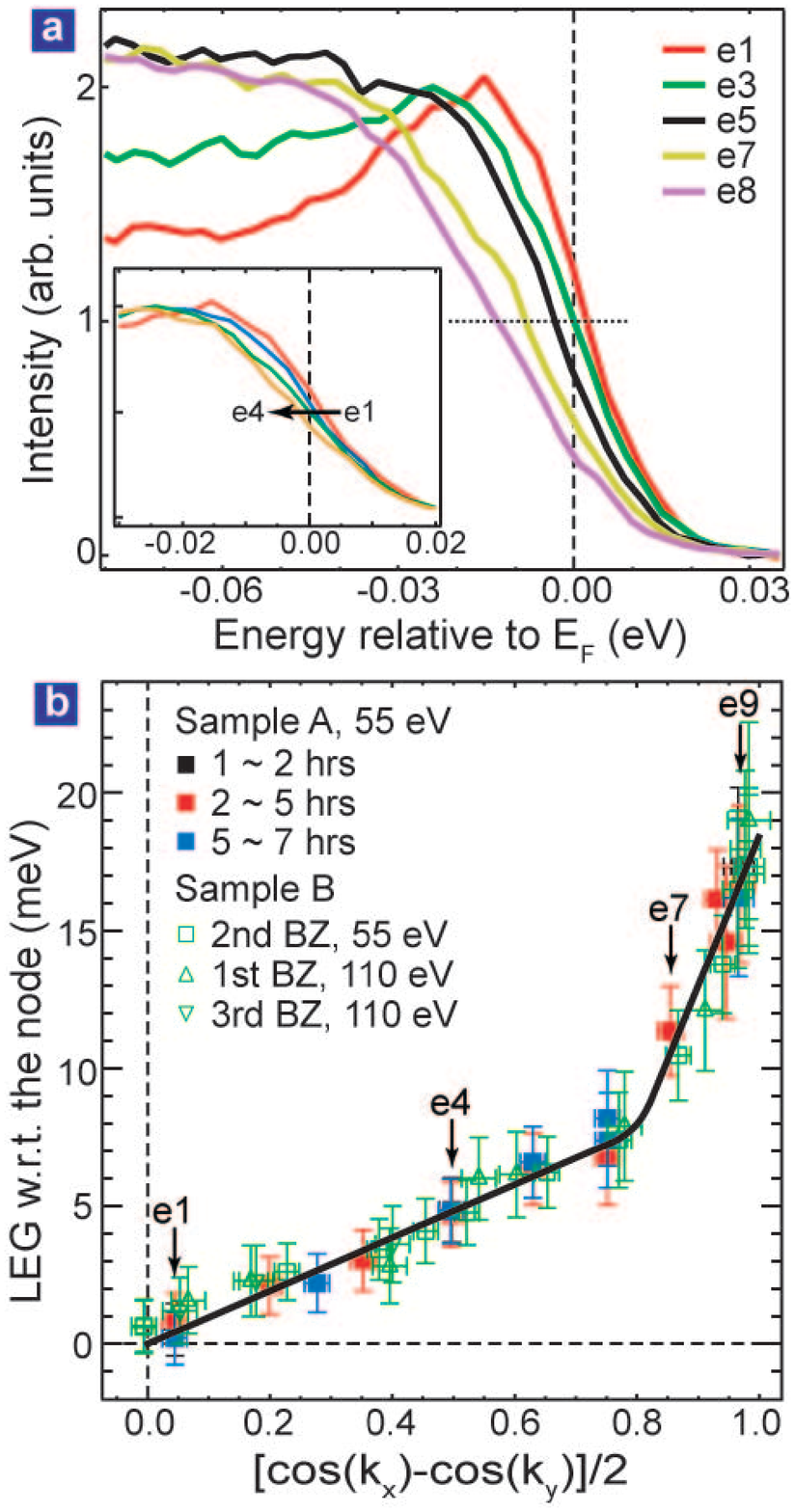}

Figure \ref{Fig. 2}
\end{figure}

\clearpage
\bigskip
\begin{figure}
\hspace*{-1cm}
\includegraphics [width=4.5 in]{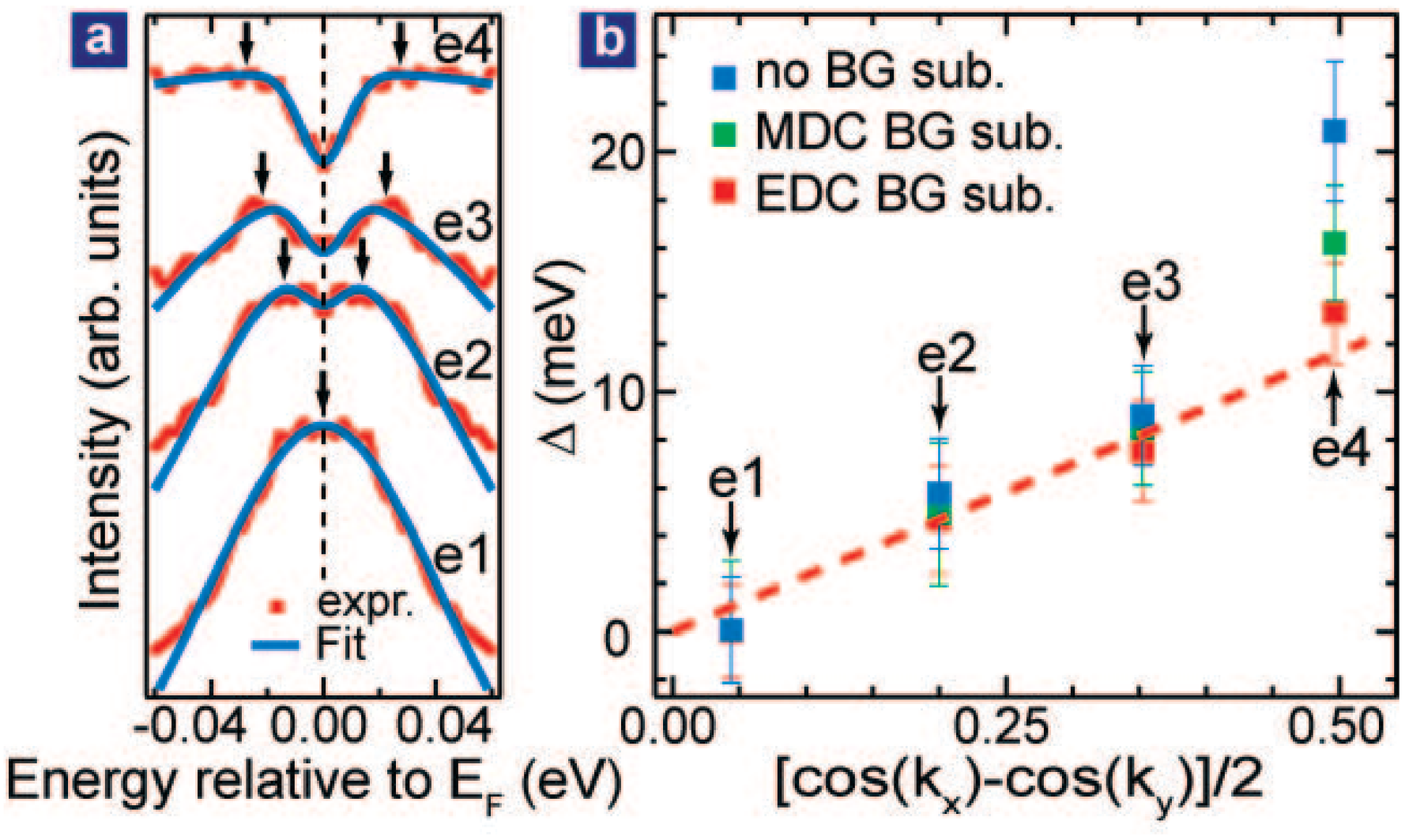}

Figure \ref{Fig. 3}
\end{figure}

\clearpage
\bigskip
\begin{figure}
\hspace*{-1cm}
\includegraphics [width=4 in]{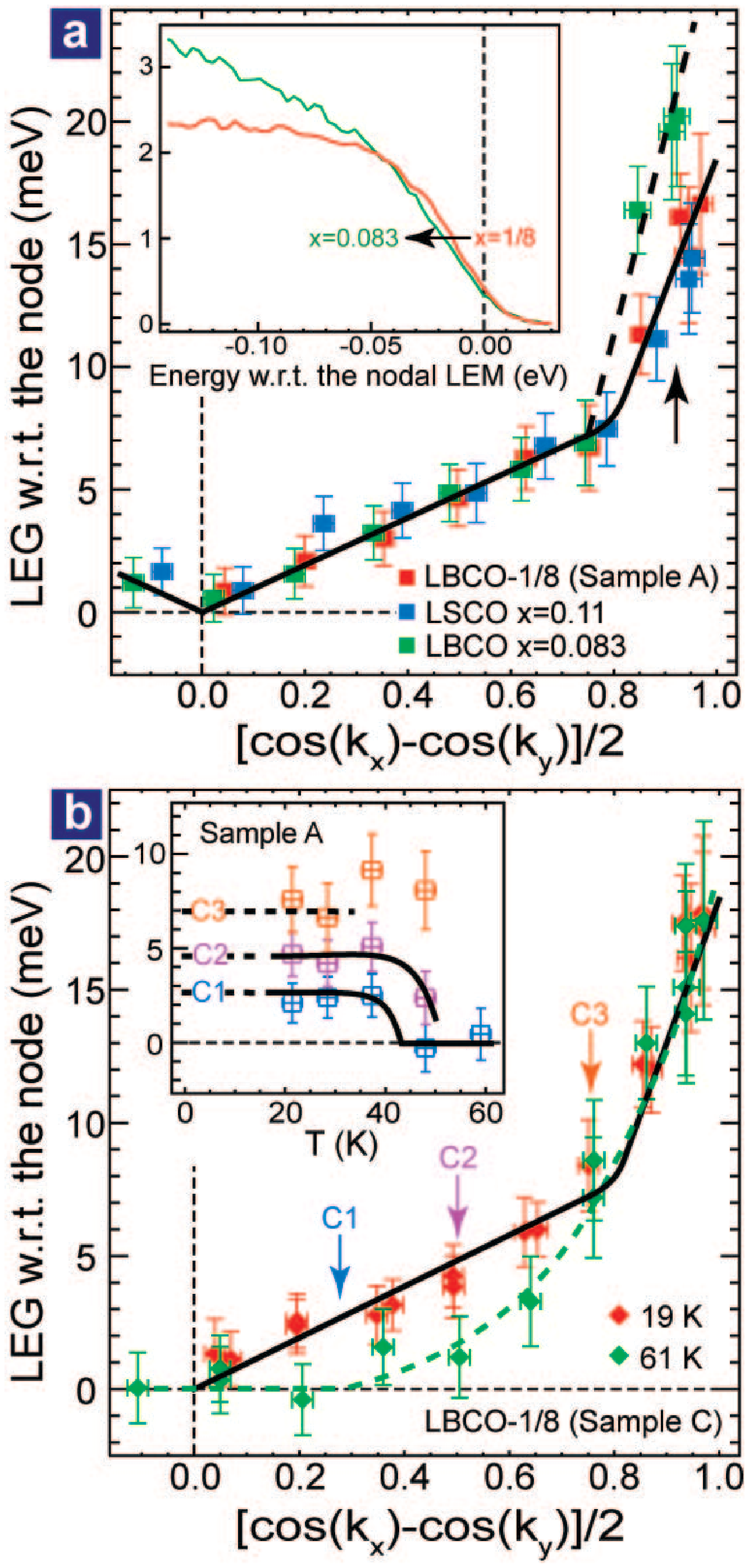}

Figure \ref{Fig. 4}
\end{figure}

\end{raggedright}

\end{document}